\documentclass{ws-ijmpd}

\usepackage{lscape}
\usepackage{color}
\newcommand{\be}{\begin{equation}}
\newcommand{\ee}{\end{equation}}
\newcommand{\bea}{\begin{eqnarray}}
\newcommand{\eea}{\end{eqnarray}}

\begin{document}

\markboth{Melia and L\'opez-Corredoira}
{AP Test with BAO}

\catchline{}{}{}{}{}
%
%%%%%%%%%%%%%%%%%%%%%%%%%%%%%%%%%%%%%%%%%%%%%%%%%%%%%%%%%%%%%%%%%%%%

\title{Model Selection using Baryon Acoustic Oscillations \\ in the Final SDSS-IV Release}

\author{F. Melia$^1$ and M. L\'opez-Corredoira$^{2,3}$}

\address{$^1$ Department of Physics, The Applied Math Program, and Department of Astronomy,
The University of Arizona, AZ 85721, USA;\\
$^2$ Instituto de Astrof\'\i sica de Canarias,\\ 
E-38205 La Laguna, Tenerife, Spain\\
$^3$ Departamento de Astrof\'\i sica, Universidad de La Laguna,\\
E-38206 La Laguna, Tenerife, Spain}

\maketitle

\begin{history}
\received{15 February 2022}
\revised{29 March 2022}
\comby{Managing Editor}
\end{history}

\begin{abstract}
The baryon acoustic oscillation (BAO) peak, seen in the cosmic matter distribution at
redshifts up to $\sim 3.5$, reflects the continued expansion of the sonic horizon first
identified in temperature anisotropies of the cosmic microwave background. The BAO
peak position can now be measured to better than $\sim 1\%$ accuracy using galaxies, and $\sim 
1.4-1.6\%$ precision with Ly-$\alpha$ forrests and the clustering of quasars. In conjunction
with the Alcock-Paczy{\'n}ski (AP) effect, which arises from the changing ratio of angular to
spatial/redshift size of (presumed) spherically-symmetric source distributions with distance,
the BAO measurement is viewed as one of the most powerful tools to use in assessing the geometry
of the Universe. In this paper, we employ five BAO peak measurements from the final release
of the Sloan Digital Sky Survey IV, at average redshifts $\langle z\rangle=0.38$, $0.51$,
$0.70$, $1.48$ and $2.33$, to carry out a direct head-to-head comparison of the standard model,
$\Lambda$CDM, and one of its principal competitors, known as the $R_{\rm h}=ct$ universe.
For completeness, we complement the AP diagnostic with a volume-averaged
distance probe that assumes a constant comoving distance scale $r_{\rm d}$. Both probes are
free of uncertain parameters, such as the Hubble constant, and are therefore ideally 
suited for this kind of model selection. We find that $R_{\rm h}=ct$ is favored 
by these measurements over the standard model based solely on the AP effect, with a likelihood 
$\sim 75\%$ versus $\sim 25\%$, while {\it Planck}-$\Lambda$CDM is favored over $R_{\rm h}=ct$
based solely on the volume-averaged distance probe, with a likelihood $\sim 80\%$ versus
$\sim 20\%$. A joint analysis using both probes produces an inconclusive outcome,
yielding comparable likelihoods to both models. We are therefore not able to confirm
with this work that the BAO data, on their own, support an accelerating Universe.
\end{abstract}

\keywords{cosmology: cosmological parameters -- cosmology: distance scale --
cosmology: observations -- quasars: general}

%PACS: 

\section{Introduction}
The early Universe contained photons, leptons and baryons coupled tightly via Compton
scattering processes and Coulomb interactions. The tussle between gravity and radiation
pressure generated oscillations and waves around overdense regions that propagated
outwards, dragging baryons and leptons along with them. Eventually, the Universe cooled
sufficiently for neutral atoms to form, after which the radiation detached from the matter,
leaving a fossilized imprint of the waves with a characteristic {\it comoving} scale which, in
the standard model ($\Lambda$CDM), is estimated to be $r_{\rm d}\sim 147$ Mpc.

This acoustic feature set the $(0.596724\pm0.00038)^\circ$ scale for the peaks seen in the
angular power spectrum of the cosmic microwave background (CMB)\cite{Planck:2020},
and continued to grow with the expansion of the Universe to influence late time galaxy
formation and evolution\cite{Peebles:1970,Sunyaev:1970,Bond:1987,Hu:1996}. 
Its presence is now seen in both Fourier and configuration space\cite{Cole:2005,Eisenstein:2005}. 
In this picture, the baryon acoustic oscillations
(BAO), as these frozen imprints are known, retained a constant comoving size after
recombination, but stretched to galaxy-cluster size along with the overall Hubble expansion.
The BAO thus represent a `standard ruler' in comoving space that one may use with measurements
at different redshifts to trace the expansion history of the Universe.

\vskip 0.2in
\noindent {\footnotesize{\bf Table 1.} Final SDSS-IV BAO Peak Data and the measured
values of $y(z)$. With
the exception of the datum at $\langle z\rangle=2.33$, all the measurements correspond 
to the consensus constraints from the combination of Fourier and configuration space BAO 
analyses, with the inclusion of a systematic error budget.}
\vskip -0.1in
\begin{table*}
\begin{center}
\begin{tabular}{cccccl}
$\langle z\rangle$ & $d_{\rm A}/r_{\rm d}$ & $d_{\rm H}/r_{\rm d}$ &
$C(d_{\rm A},d_{\rm H})$ & $y(z)$ & Reference\\
\hline\hline
\\
0.38 & $7.41\pm0.12$ & $25.00\pm0.76$ & $-0.29$ & $1.076\pm0.041$ & \cite{Alam:2017,Cuesta:2016} \\
0.51 & $8.85\pm0.14$ & $22.33\pm0.58$ & $-0.50$ & $1.173\pm0.043$ & \cite{Alam:2017} \\
0.70 & $10.51\pm0.19$ & $19.33\pm0.53$ & $-0.50$ & $1.320\pm0.052$ & \cite{Gil-Marin:2020,Bautista:2021} \\
1.48 & $12.38\pm0.32$ & $13.26\pm0.55$ & $-0.50$ & $1.564\pm0.092$ & \cite{Neveux:2020,Hou:2021} \\
2.33 & $11.26\pm0.33$ & $8.99\pm0.19$ & $-0.40$ & $1.790\pm0.076$ & \cite{duMas:2021} \\ \\
\hline\hline
\end{tabular}
\end{center}
\end{table*}

Over the past two decades, the Sloan Digital Sky Survey (SDSS)\cite{York:2000} has
constructed several diverse spectroscopic catalogs of galaxies and quasars one may use
to map the large-scale structure of the Universe. The last of these programs, called
eBOSS (extended Baryon Oscillation Spectroscopic Survey)\cite{Dawson:2016}, in the
fourth phase of SDSS (referred to as SDSS-IV)\cite{Blanton:2017}, includes four main
tracers: luminous red galaxies, emission-line galaxies, quasars, and a separate
high-redshift quasar sample for the study of the Ly-$\alpha$ forest. The eBOSS program
came to an end on March 1st, 2019.

In this paper, we shall adopt source bins from three of these
catalogs with effective (i.e., `average') redshifts ranging from 0.38 to 2.33.
The BAO detection using quasars as tracers\cite{Neveux:2020,Hou:2021} at $0.8\lesssim z\lesssim 2.2$ 
bridges the gap between the lower redshift SDSS galaxy surveys\cite{Alam:2017,Alam:2021}
at $z\lesssim 0.8$ and the Ly-$\alpha$ forest\cite{Bautista:2017,duMas:2021} at $z\gtrsim 2.2$.

A principal motivation for the analysis we carry out here, based on the use of these
BAO measurements with the Alcock-Paczy\'nski effect described below\cite{Alcock:1979}, 
complemented by a volume-averaged distance probe that assumes a constant
value of $r_{\rm d}$, is the significant improvement achieved recently in mitigating the 
contamination due to so-called redshift-space distortions. These are due to peculiar 
velocities induced by gravitational influences within the clusters 
themselves\cite{Kaiser:1987,Matsubara:1996,Hamilton:1998}, 
but methods have been developed to break the degeneracy based on the notion that 
the distortions affect primarily the amplitude of the BAO peak, while their impact 
on its position is negligible\cite{Font-Ribera:2014}. And it is the BAO peak's
position that determines the standard ruler used for the Alcock-Paczy\'nski 
and volume-averaged distance tests.

This outcome is achieved via reconstruction techniques\cite{Eisenstein:2007,Padmanabhan:2012} 
that enhance the quality of the galaxy two-point correlation function. In
addition, the newer Ly-$\alpha$ and quasar auto- and cross-correlation functions may not be
as precise as the BAO peak position measured with galaxies (i.e., $\sim 1\%$), but are 
still sufficiently accurate to provide a BAO peak position with an accuracy better than 
$\sim 1.4\%$ and $\sim 1.6\%$, respectively\cite{Bautista:2018}. One should note that 
all of these measurements 
use a template for the correlation function drawn from the concordance model, which could 
create a problem when one uses alternative cosmologies. But the actual shape of the BAO peak 
barely influences the determination of its centroid position, both parallel to the line-of-sight 
and in the perpendicular direction, as long as its FWHM is very narrow. For the most recent 
measurements we use in this analysis, the overall redshift distortions produce systematic 
errors of order $\lesssim 0.5\%$ in the angular-diameter ($d_{\rm A}[z]$) and comoving 
($d_{\rm com}[z]$) distances---an impressive feat, given that statistical errors as small 
as $\sim 4\%$ are now achievable (see refs.~\cite{Hou:2021,Alam:2021}; see also
refs.~\cite{Gil-Marin:2020,Neveux:2020,Bautista:2021}).

In the next section, we describe how the BAO peak positions can be used to assess
the geometry of the Universe in a model-independent way, even without knowing a
precise value for the Hubble constant, $H_0$. Our use of this approach in an
earlier application based on older data\cite{Melia:2017} produced
a very intriguing outcome, notably that the model preferred by the BAO data is not
necessarily $\Lambda$CDM. A principal goal of such studies is to probe the hypothesized
late-time acceleration of the Universe's expansion. In this paper, we carry out
this analysis using the updated, final SDSS-IV release to test three competing cosmologies:
(i) {\it Planck}-$\Lambda$CDM\cite{Planck:2020}; (ii) $\Lambda$CDM with its
matter-density ($\Omega_{\rm m}$) as a free, optimizable parameter, and (iii) the
$R_{\rm h}=ct$ universe\cite{Melia:2012,Melia:2020}, whose viability has
been established previously using over 27 different kinds of data. The critical
difference between a cosmology based on $\Lambda$CDM versus one characterized by
$R_{\rm h}=ct$ is that, while the former accelerates, the latter always expands
at a constant rate.

\section{BAO Probes}
\subsection{The Alcock-Paczy\'nski Test}
The measurements of $d_{\rm A}(z)/r_{\rm d}$ and $d_{\rm com}(z)/r_{\rm d}$ described above are
ideally suited to the Alcock-Paczy\'nski\cite{Alcock:1979} test we shall now describe.
Critically, the actual value
of $r_{\rm d}$ is not required, nor is the Hubble constant, $H_0$, whose measurements at low and
high redshifts now appear to be inconsistent with each other at a $4.4\sigma$ level of
significance. The Hubble constant characterizes the current expansion rate of the Universe
and determines its {\it absolute} distance scale, but as the accuracy with which it is measured
continues to improve, its value\cite{Planck:2020} ($67.4\pm0.5$ km $\rm s^{-1}$ $\rm Mpc^{-1}$) 
inferred from the CMB in the context of flat $\Lambda$CDM contrasts with that 
($74.03\pm1.42$ km $\rm s^{-1}$ $\rm Mpc^{-1}$) based on local Type Ia supernovae calibrated 
with the Cepheid distance ladder\cite{Riess:2019}. A cosmological probe that
avoids having to use $H_0$---such as the Alcock-Paczy\'nski test we employ
here---therefore has a significant advantage over others that require the measurement
of absolute distances.

Assuming we have a spherically-symmetric distribution of objects (or, equivalently,
a standard ruler whose length in the comoving frame is independent of orientation with respect
to our line-of-sight), with radius 
\begin{eqnarray}
L_\parallel&=&\Delta z\frac{d}{dz}d_{\rm com}(z)\label{eq:Lpar}\nonumber\\
\null&=&\Delta z\,d_{\rm H}(z)
\end{eqnarray}
along the line-of-sight and
\begin{equation}
L_\perp =\Delta \theta (1+z) d_{\rm A}(z)\label{eq:Lperp}
\end{equation}
in the perpendicular direction, the ratio
\begin{equation}
y(z)\equiv \frac{\Delta z}{z\Delta \theta }\frac{L_\perp}{L_\parallel }\label{eq:yz}
\end{equation}
depends only on the cosmological comoving distance, $d_{\rm com}(z)$, and the
angular-diameter distance, $d_{\rm A}(z)$ and, very importantly, is independent of any
source evolution. The quantity $d_{\rm H}(z)$ is the Hubble distance
defined in Equation~(7) below.

These two distances are simply related according to the equation\cite{Melia:2012,Melia:2020}
\begin{equation}
d_{\rm com}^R(z)=(1+z)\,d_A^R(z)=\frac{c}{H_0}\ln (1+z)\;,\label{eq:dcomR}
\end{equation}
in the case of $R_{\rm h}=ct$ and, more generally, via the parametrized formulation
\begin{equation}
d_{\rm com}^\Lambda(z)=(1+z)\,d_A^\Lambda(z)=\frac{c}{H_0}
\int _0^z\frac{du}{\sqrt{\Omega _{\rm m}(1+u)^3+\Omega_\Lambda }}\;,\label{eq:da_wCDM}
\end{equation}
in $\Lambda$CDM. This expression assumes spatial flatness and ignores the infinitesimal
contribution from radiation in the redshift range of interest. It also employs the
conventional definition of a scaled density, $\Omega_i$, representing the energy density
of species $i$ normalized to today's critical density, $\rho_{\rm c}\equiv 3c^2H_0^2/8\pi G$.

\begin{figure*}[t]
\centering
\includegraphics[angle=0,scale=0.52]{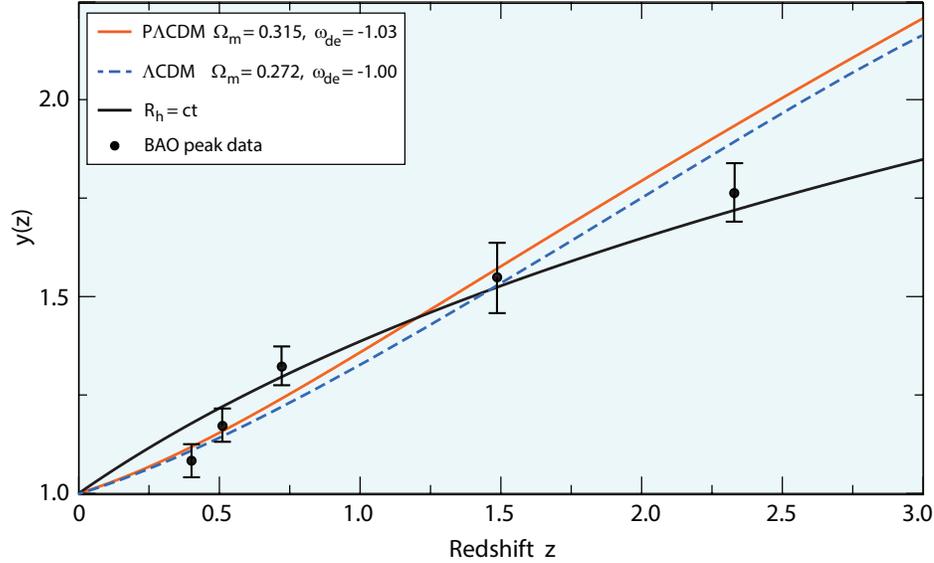}
\caption{Ratio of observed angular size to radial/redshift size, $y(z)$, versus
redshift, $z$, inferred from five BAO peak measurements at $\langle z\rangle = 0.38$, $0.51$,
$0.70$, $1.48$ and $2.33$. Note, however, that the three lowest-redshift data points
are not completely independent (see text). The data are compared to the predictions of 3 models:
(i) (red solid) {\it Planck}-$\Lambda$CDM, with (pre-optimized) parameters $\Omega_{\rm m}=0.315$
and $w_{\rm de}=-1.03$, (ii) (blue dash) a best-fit $\Lambda$CDM
cosmology using $\Omega_{\rm m}$ as a free parameter (with optimized value $0.292\pm 0.038$,
and a fixed $w_{\rm de}=-1.00$), and (iii) (black solid) the $R_{\rm h}=ct$ universe, whose $y(z)$
function is completely free of any parameters.}\label{fig1}
\end{figure*}

It is therefore rather trivial to see that
\begin{equation}
y(z)={(1+z)\over z}{d_{\rm A}(z)\over d_{\rm H}(z)}\;,\label{eq:yprefinal}
\end{equation}
where
\begin{equation}
d_{\rm H}\equiv {c\over H(z)}\;,\label{eq:dH}
\end{equation}
and
\begin{equation}
H^R(z)\equiv H_0(1+z)\;,\label{eq:HR}
\end{equation}
for $R_{\rm h}=ct$, while
\begin{equation}
H^\Lambda(z)\equiv H_0\sqrt{\Omega _{\rm m}(1+z)^3+\Omega_\Lambda}\;,\label{eq:HL}
\end{equation}
for the standard model. Clearly, both $H_0$ and $r_{\rm d}$ cancel completely in $y(z)$
when we write
\begin{equation}
y(z)={(1+z)\over z}{d_{\rm A}(z)/r_{\rm d}\over d_{\rm H}(z)/r_{\rm d}}\;,\label{eq:yfinal}
\end{equation}
allowing us to utilize the data in Table~1 in a model-independent fashion.
This compilation includes five measurements of $d_{\rm A}/r_{\rm d}$ and
$d_{\rm H}/r_{\rm d}$, their individual errors, and the correlation coefficient between
$d_{\rm A}$ and $d_{\rm H}$, labeled $C(d_{\rm A},d_{\rm H})$, in column 4. The uncertainty
in $y(z)$ is estimated according to the error propagation equation
\begin{equation}
\sigma_y^2 = \left(y{\sigma_{d_{\rm A}}\over d_{\rm A}}\right)^2+
\left(y{\sigma_{d_{\rm H}}\over d_{\rm H}}\right)^2-2y^2{\sigma_{d_{\rm A}.d_{\rm H}}\over
d_{\rm A}\,d_{\rm H}}\;,
\end{equation}
where
\begin{equation}
\sigma_{d_{\rm A}.d_{\rm H}}\equiv C(d_{\rm A},d_{\rm H})\,\sigma_{d_{\rm A}}\sigma_{d_{\rm H}}\;.
\end{equation}

\begin{figure}[t]
\centering
\includegraphics[angle=0,scale=0.8]{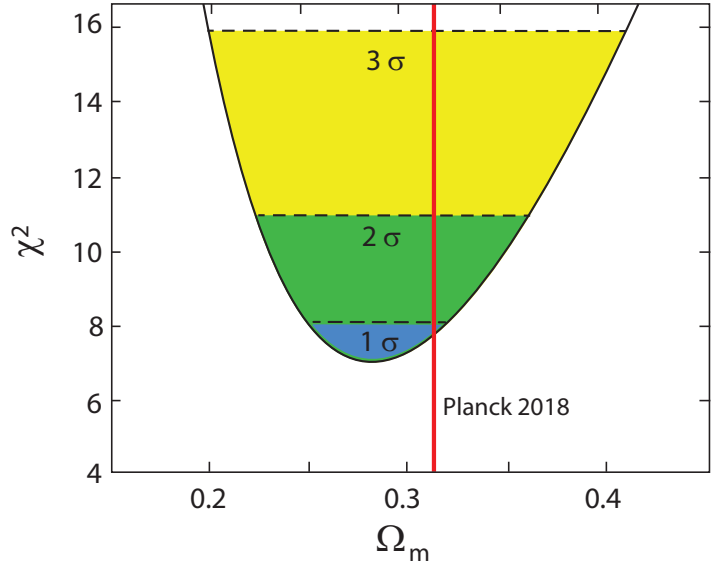}
\caption{Optimization of the matter density parameter $\Omega_{\rm m}$ for the $\Lambda$CDM
model in figure~\ref{fig1}, with the $1\sigma=0.04$, $2\sigma$ and $3\sigma$ confidence levels
in a Gaussian distribution, i.e., $68.3\%$ (cyan), $95.4\%$ (green) and $99.73\%$ (yellow). By
comparison, the {\it Planck}-$\Lambda$CDM (concordance) model corresponds to the optimized
parameter value $\Omega_{\rm m}=0.315\pm 0.007$, which is consistent
with our best fit value ($\Omega_{\rm m}=0.272\pm0.041$) to within $1\sigma$.}\label{fig2}
\end{figure}

Figure~\ref{fig1} shows the function $y(z)$ measured with the five BAO data points summarized
in Table~1, along with the predictions of three models: (i) (solid black) $R_{\rm h}=ct$;
(ii) (solid red) {\it Planck}-$\Lambda$CDM (with pre-optimized parameters $\Omega_{\rm m}=0.315$,
$\Omega_{\rm de}=1-\Omega_{\rm m}$, and a dark-energy equation-of-state parameter\cite{Planck:2020}
$w_{\rm de}=-1.003$); and (iii) (dashed blue) a more generic
$\Lambda$CDM model with a free $\Omega_{\rm m}$ variable. In this case, the fit is
optimized with a matter density $\Omega_{\rm m}=0.272\pm0.041$ (see also fig.~\ref{fig2}), which
is consistent to within $1\sigma$ with the {\it Planck} value.

The quality of each fit may be assessed via its corresponding reduced $\chi^2$ value, i.e.,
$\chi^2_{\rm dof}\equiv \chi^2/[N-\nu]$, where $N$ is the number of independent
data points and $\nu$ is the number of free parameters in the model. The bins used to
obtain the two lowest-redshift measurements in this set overlap somewhat and are therefore
not completely independent. They provide a net redshift coverage of $0.4$, but their bins
span a total range of $0.5$. We therefore estimate the `effective' number of independent
points as $2\times (0.4/0.5)=1.6$ for this subset, yielding a total number $N=4.6$ of
independent data points. In estimating the $\chi^2$ for each model, we thus use
\begin{equation}
\chi^2=\sum_{i=1}^5f_i{[y_{\rm th}(z_i)-y(z_i)]^2\over\sigma_{y(z_i)}^2}\;,\label{eq:chi2}
\end{equation}
where $y_{\rm th}$ is the theoretical value, $y(z_i)$ is the observed quantity, and
$f_i=0.8$ for $i=1$--$2$, and $f_i=1$ for points $3$ to $5$.

\vskip 0.2in
\noindent {\footnotesize{\bf Table 2.} Model comparison using the effective number, 
$N=4.6$, of BAO peak data points, based on (i) the reduced $\chi^2_{\rm dof}\equiv\chi^2/(N-\nu)$, 
with $\nu=$ number of free parameters; (ii) optimized parameters (if any); (iii) AIC; and (iv) KIC
(see \S~3).}
\vskip -0.1in
\begin{table*}
\begin{center}
\begin{tabular}{lccrr}
Model & $\chi^2_{\rm dof}$ & Parameter & AIC & KIC \\ \hline\hline
\\
$\Lambda$CDM & \quad$\,$1.84 \quad\qquad & \quad $\Omega_{\rm m}=
0.272\pm0.041$ \quad\qquad & 8.61 & 9.61 \\
{\it Planck}-$\Lambda$CDM \quad\quad & 1.74 & -- & 8.00 & 
8.00 \\
$R_{\rm h}=ct$ & 1.43 & -- & 6.58 & 6.58 \\ \\
\hline\hline
\end{tabular}
\end{center}
\end{table*}

In the Appendix, we gauge the impact of this approach using a covariance matrix
available online\footnote{https://svn.sdss.org/public/data/eboss/DR16cosmo/tags/v1-0-1/likelihoods/BAO-only/} 
for $d_{\rm com}$ and $d_{\rm H}$ in the binnings at $z=0.38$ and $0.51$, along with an estimate of 
the number of galaxies in the overlapping redshift regions based on the overlap redshift intervals. 
A comparison of the results summarized in Tables~5 and 6 with those shown in Tables~2 and 4 suggests 
that these two procedures, using Equations~(\ref{eq:chi2}) and (\ref{eq:chiACA}), produce only minor 
relative changes to the outcome of our model comparison. The model prioritization, however, remains 
unchanged.

Very interestingly, this Alcock-Paczy\'nski test, based on BAO measurements, provides
an optimized matter density parameter, $\Omega_{\rm m}=0.272\pm0.041$ (or
$0.276\pm0.041$ if we instead use Eq.~\ref{eq:chiACA}), fully consistent (to within
$1\sigma$) with its {\it Planck} value. One can see this quite easily by comparing the two curves in
Figure~\ref{fig1}, which show that the optimized fit to $y(z)$ is very close to that of the
{\it Planck}-$\Lambda$CDM prediction (see also Fig.~\ref{fig2}). Quite tellingly, however,
this figure also demonstrates why $R_{\rm h}=ct$ is favored by the BAO data and the
AP diagnostic over both {\it Planck}-$\Lambda$CDM and the more generic version of $\Lambda$CDM 
with a variable $\Omega_{\rm m}$ (see Table~2). It should also be stressed that $R_{\rm h}=ct$
has no free parameters at all for this comparison with the data. In other words, these
cosmologies are not all nested, so a statistically fair comparison needs to take into
account the different numbers of free parameters.  We shall address this issue in 
\S~3 below.

\subsection{Volume-averaged distance probe}
The AP effect offers a clean cosmological test that requires minimal assumptions,
but it ignores another possibly important piece of information, i.e., that the
comoving scale $r_{\rm d}$ characterizing the BAO signal should be the same at all 
redshifts---at least in the context of $\Lambda$CDM. The volume-averaged distance
probe is thus not as clean as the AP effect, which is based solely on the geometry,
whereas the former must make some assumption concerning the evolution (or not) of
the comoving scale $r_{\rm d}$. Nevertheless, we shall here assume for simplicity
that $r_{\rm d}$ is indeed constant in the models we examine. 

\vskip 0.2in
\noindent {\footnotesize{\bf Table 3.} Final SDSS-IV BAO Peak Data and the measured values of $x(z)$.} 
\vskip -0.1in
\begin{table*}
\begin{center}
\begin{tabular}{cccccl}
$\langle z\rangle$ & $d_{\rm A}/r_{\rm d}$ & $d_{\rm H}/r_{\rm d}$ &
$C(d_{\rm A},d_{\rm H})$ & $x(z)$ & Reference\\
\hline\hline
\\
0.38 & $7.41\pm0.12$ & $25.00\pm0.76$ & $-0.29$ & $1.000\pm0.022$ & \cite{Alam:2017,Cuesta:2016} \\
0.51 & $8.85\pm0.14$ & $22.33\pm0.58$ & $-0.50$ & $1.269\pm0.042$ & \cite{Alam:2017} \\
0.70 & $10.51\pm0.19$ & $19.33\pm0.53$ & $-0.50$ & $1.632\pm0.084$ & \cite{Gil-Marin:2020,Bautista:2021} \\
1.48 & $12.38\pm0.32$ & $13.26\pm0.55$ & $-0.50$ & $2.650\pm0.120$ & \cite{Neveux:2020,Hou:2021} \\
2.33 & $11.26\pm0.33$ & $8.99\pm0.19$ & $-0.40$ & $3.095\pm0.116$ & \cite{duMas:2021} \\ \\
\hline\hline
\end{tabular}
\end{center}
\end{table*}

A complete analysis of BAO measurements often includes a constraint on the ratio 
$D_v(z)/r_{\rm d}$, where the volume-averaged distance $D_v(z)$ is defined as
\begin{eqnarray}
D_v(z) &\equiv& \left\{z\,d_{\rm com}(z)^2 d_{\rm H}(z)\right\}^{1/3}\nonumber\\
\null&=&\left\{z(1+z)^2d_{\rm A}(z)^2d_{\rm H}(z)\right\}^{1/3}\;.\label{eq:Dv}
\end{eqnarray}

Together, the AP and $D_v$ diagnostics constitute a more complete representation 
of the full information content of the BAO signal---assuming, of course, that $r_{\rm d}$
is indeed constant in all the models being tested. Using the volume-averaged distance
$D_v$ on its own, however, would require a model-dependent prediction of $r_{\rm d}$, 
which would complicate the analysis, and possibly weaken the model comparison.
Instead, we shall use the ratio
\begin{equation}
x(z) \equiv {D_v(z)/r_{\rm d} \over  D_v(z_{\rm p})/r_{\rm d}}\;,\label{eq:xz}
\end{equation}
using a `pivot' redshift, $z_{\rm p}$, for each of the remaining four measurements.
We choose the pivot point $\langle z\rangle=0.38$. In the case of $y(z)$, the values 
of $r_{\rm d}$ and $H_0$ completely cancel out. The volume-averaged distance probe, 
$x(z)$, is also independent of both $r_{\rm d}$ and $H_0$, though the required use 
of the ratio in Equation~(\ref{eq:xz}) then leaves us with only four data points, 
which are shown in Table~3. The uncertainty in $x(z)$ is estimated according to 
the error propagation equation
\begin{equation}
\sigma_x^2 = \left(x{2\sigma_{d_{\rm A}}\over 3d_{\rm A}}\right)^2+
\left(x{\sigma_{d_{\rm H}}\over 3d_{\rm H}}\right)^2-2x^2{\sigma_{d_{\rm A}.d_{\rm H}}\over
d_{\rm A}\,d_{\rm H}}+\left(x{\sigma_{\rm piv}\over x(0.38)}\right)^2\;,\label{eq:sigx}
\end{equation}
where
\begin{equation}
\sigma_{d_{\rm A}.d_{\rm H}}\equiv C(d_{\rm A},d_{\rm H})\,\sigma_{d_{\rm A}}\sigma_{d_{\rm H}}\;,
\end{equation}
and $\sigma_{\rm piv}$ is the error in the pivot value $x(0.38)$. (Note that the last
term in Eq.~\ref{eq:sigx} is used only for the non-pivot points.)

\begin{figure}[t]
\centering
\includegraphics[angle=0,scale=0.63]{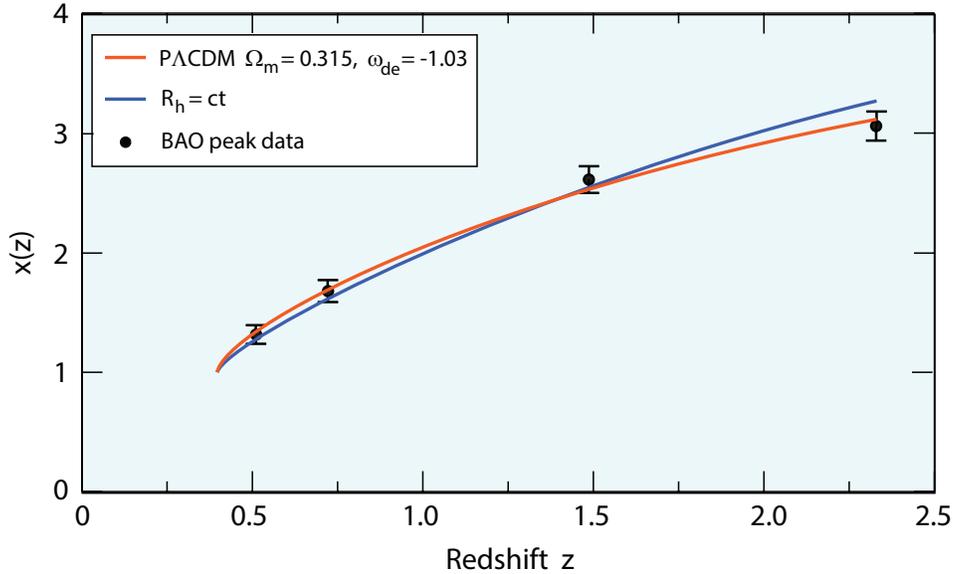}
\caption{The volume-averaged quantity $x(z)$ (see Eq.~\ref{eq:xz}), versus redshift,
$z$, inferred from the four BAO peak measurements at $\langle z\rangle=0.51$, $0.70$,
$1.48$ and $2.33$ (see Table~3). The datum at $\langle z\rangle=0.38$ is not included
because it is used as the `pivot' point (see Eq.~\ref{eq:xz} and Table~3). The data are 
compared to the predictions of 2 models: (i) (red solid) {\it Planck}-$\Lambda$CDM, with 
(pre-optimized) parameters $\Omega_{\rm m}=0.315$ and $w_{\rm de}=-1.03$, and (ii) 
(blue solid) the $R_{\rm h}=ct$ universe, whose $x(z)$ function is completely free 
of any parameters.}\label{fig3}
\end{figure}

The volume-averaged quantity $x(z)$ is shown in figure~\ref{fig3} for the four measured
BAO data points summarized in Table~3, along with the predictions of: (i) 
(solid blue) $R_{\rm h}=ct$, and (ii) (solid red) {\it Planck}-$\Lambda$CDM (with 
pre-optimized parameters $\Omega_{\rm m}=0.315$, $\Omega_{\rm de}=1-\Omega_{\rm m}$, 
and a dark-energy equation-of-state parameter\cite{Planck:2020} $w_{\rm de}=-1.003$). 
The curve corresponding to the optimization of $\Lambda$CDM with one free parameter 
is very similar to these two curves, so we omit it for the sake of clarity.

In this case, all four of the data points are independent of each other. The $\chi^2$ values 
for the three models are $3.216$ (yielding $\chi^2_{\rm dof}=0.804$) for $R_{\rm h}=ct$, 
$0.421$ (yielding $\chi^2_{\rm dof}=0.105$) for {\it Planck}-$\Lambda$CDM, and
$0.409$ (yielding $\chi^2_{\rm dof}=0.105$) for $\Lambda$CDM, with an optimized
matter density parameter $\Omega_{\rm m}=0.308\pm0.040$. Based solely on their values
of $\chi^2$, all three models fit the volume-averaged distances very well. As we shall 
see in \S~3, however, the information criteria somewhat favor both {\it Planck}-$\Lambda$CDM 
and $\Lambda$CDM over $R_{\rm h}=ct$. We shall quantitatively assess the impact of these 
measurements on the overall model selection in the next section.

\section{Model Selection}
To compare models with a different number of free parameters, constituting
an uneven flexibility in fitting the data, it is now common in cosmology to
carry out model selection via the use of information criteria. These include
the Akaike Information Criterion\cite{Akaike:1974,Liddle:2007,Burnham:2002,MeliaMaier:2013},
\begin{equation}
{\rm AIC}\equiv\chi^2+2\nu\label{eq:AIC}\;,
\end{equation}
and the Kullback Information Criterion\cite{Cavanaugh:2004},
\begin{equation}
{\rm KIC}\equiv\chi^2+3\nu\label{eq:KIC}\;.
\end{equation}

\vskip 0.2in
\noindent {\footnotesize{\bf Table 4.} Head-to-head model comparisons 
using the AIC and KIC values in Table~2, and those corresponding to the fits shown
in Fig.~\ref{fig3}.}
\vskip -0.1in
\begin{table*}
\begin{center}
\begin{tabular}{lcccc}
Model Comparison & $\Delta {\rm AIC}$ & AIC & $\Delta {\rm KIC}$ & KIC \\
\hline\hline
\\
$\underline{\rm AP\;effect\;only}$&&&& \\
{\it Planck}-$\Lambda$CDM vs. $R_{\rm h}=ct$\qquad & 1.42 & 
$33\%$ vs. $67\%$ & 1.42 & $33\%$ vs. $67\%$ \\
$\Lambda$CDM vs. $R_{\rm h}=ct$ & 2.03 & $27\%$ vs. $73\%$ & 
3.03 & $18\%$ vs. $82\%$ \\ \\
$\underline{\rm Volume\;averaged\;distance\;only}$&&&& \\
{\it Planck}-$\Lambda$CDM vs. $R_{\rm h}=ct$\qquad & -2.79\; &
$80\%$ vs. $20\%$ & -2.79\; & $80\%$ vs. $20\%$ \\ 
$\Lambda$CDM vs. $R_{\rm h}=ct$ & -0.81\; & $60\%$ vs. $40\%$ &
0.19 & $48\%$ vs. $52\%$ \\ \\
$\underline{\rm Combined\;test}$&&&& \\
{\it Planck}-$\Lambda$CDM vs. $R_{\rm h}=ct$\qquad & -1.37\; &
$66\%$ vs. $34\%$ & -1.37\; & $66\%$ vs. $34\%$ \\
$\Lambda$CDM vs. $R_{\rm h}=ct$ & -0.37\; & $55\%$ vs. $45\%$ &
1.37 & $33\%$ vs. $67\%$ \\ \\
\hline\hline
\end{tabular}
\end{center}
\end{table*}

A third variant, known as the Bayes Information Criterion\cite{Schwarz:1978}, 
is an asymptotic ($N\rightarrow\infty$) approximation to the
outcome of a conventional Bayesian inference procedure for deciding between
models. This criterion, however, is reliable only when the number of data
points $N$ is large (i.e., $\gg 20$). For the analysis in this paper,
we have $N=4.8$ using the AP probe, and $4$ for the volume-averaged measure
of distance, $x(z)$, so we restrict our attention solely to the AIC and KIC.

These information criteria provide a consistent way to assess which model
is favored by the data. For model $\mathcal{M}_\alpha$, the unnormalized confidence
of it being `true' is the Akaike weight $\exp(-{\rm AIC}_\alpha/2)$. In a head-to-head
comparison between two competing models, its relative likelihood of being the correct
choice is therefore
\begin{equation}
P(\mathcal{M}_\alpha)= \frac{\exp(-{\rm AIC}_\alpha/2)}
{\exp(-{\rm AIC}_1/2)+\exp(-{\rm AIC}_2/2)}\;.\label{eq:Prob}
\end{equation}
\vskip 0.04in
\noindent An analogous expression is used for KIC. In Table~4, we show 
the model comparison likelihoods for $y(z)$ only, $x(z)$ only, and finally for the 
joint test using {\it both} the AP and volume-averaged distance probes.

\begin{figure}[t]
\centering
\includegraphics[angle=0,scale=1.0]{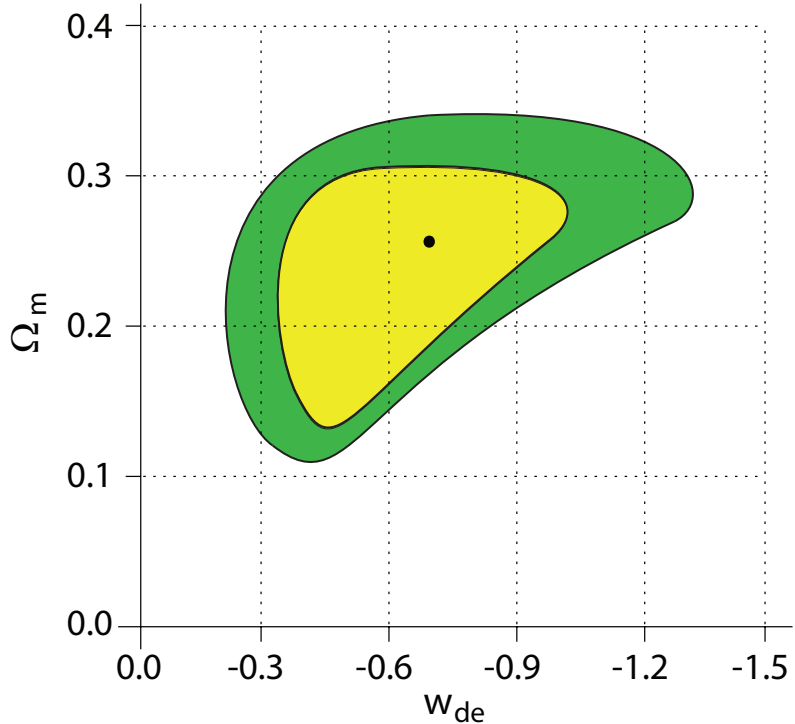}
\caption{Confidence level contours as a function of $\Omega_{\rm m}$ and
$w_{\rm de}$ for the $w$CDM model with two free parameters optimized via the
combined AP and volume-averaged distance probes. The two levels are equivalent
to $1-2\sigma$ in a Gaussian distribution (i.e., $68.3\%$ C.L., yellow;
$98.4\%$ C.L., green). The best fit parameters, indicated by the black dot,
are $\Omega_{\rm m}=0.256\pm0.081$ and $w_{\rm de}=-0.68\pm0.31$. Both differ from
the {\it Planck} optimization adopted for the {\it Planck}-$\Lambda$CDM
model included in Tables~2 and 4, but are nevertheless consistent with
those values to within $1\sigma$.}\label{fig4}
\end{figure}

The reduced $\chi^2_{\rm dof}$ values in Table~2 and the percentage 
likelihoods in Table~4 show that the application of the Alcock-Paczy\'nski test to 
the BAO peak data favors the $R_{\rm h}=ct$ universe over both {\it Planck}-$\Lambda$CDM 
and flat $\Lambda$CDM with an adjustable $\Omega_{\rm m}$ parameter. The generic version 
of $\Lambda$CDM optimized to fit these BAO data is fully consistent with the {\it Planck} 
model. 

When the BAO data are used with the volume-averaged distance probe, $D_v$, however,
the likelihoods are reversed. Based solely on their $\chi^2_{\rm dof}$ values, both
$R_{\rm h}=ct$ and {\it Planck}-$\Lambda$CDM fit the $x(z)$ data comparably well. The 
information criteria, however, show a tangible preference for the concordance model 
over $R_{\rm h}=ct$, with likelihoods $\sim 80\%$ versus $\sim 20\%$. But the outcome
based on the volume-averaged distance probe is inconclusive when comparing $\Lambda$CDM
and $R_{\rm h}=ct$, yielding comparable likelihoods for both models.

The most compelling model selection ought to be the joint analysis using both the
AP and volume-averaged distance probes, whose outcome is also summarized in Table~4.
As one can see, however, this procedure yields comparable likelihoods for all three
models, suggesting that the BAO data used in conjunction with the AP and $D_v$ tests
do not strongly support the idea that the Universe is currently accelerating.

Finally, we address the possibility that restricting $\Lambda$CDM to just one free 
parameter may be inhibiting its ability to fit the BAO data. We have therefore also 
carried out the joint analysis for model selection between $R_{\rm h}=ct$ and $w$CDM, 
in which we relax the requirement that dark energy be a cosmological constant. In 
this case, the standard model has two free parameters: $\Omega_{\rm m}$ and the
dark-energy equation of state variable $w_{\rm de}$. 

The results of this comparison are summarized in Figure~\ref{fig4}, which shows
the confidence level contours as a function of $\Omega_{\rm m}$ and $w_{\rm de}$.
Based on the joint AP-$D_{v}$ analysis, the $w$CDM model best fits the BAO data 
with the optimized parameters $\Omega_{\rm m}=0.256\pm0.081$ and $w_{\rm de}=-0.68\pm0.31$.
These are noticeably different from their {\it Planck} counterparts, but are consistent
with them to within 1-$\sigma$. In this case, the AIC assigns the likelihoods $55\%$ versus 
$45\%$ favoring $w$CDM over $R_{\rm h}=ct$, while the KIC favors $R_{\rm h}=ct$ over
$w$CDM with likelihoods $69\%$ versus $31\%$. Notice also that the likelihoods for
$w$CDM are virtually indistinguishable from those assigned to $\Lambda$CDM. This
happens because the greater flexibility afforded by the extra free parameter in
$w$CDM is mitigated by the additional penalty assigned to this model by the information
criteria. In other words, our joint analysis using both the AP and $D_v$ probes
does not provide a clear indication of which model is favored by the BAO data.

\section{Discussion and Conclusion}
The analysis we have carried out is based exclusively on the Alcock-Paczy\'nski 
effect, formalized in Equation~(\ref{eq:yfinal}), and the volume-averaged distance measure,
defined in Equation~(\ref{eq:xz}). The data corresponding to the BAO scale constitute independent 
measurements of the comoving distance, $d_{\rm com}(z)/r_{\rm d}$, and the Hubble distance, 
$d_{\rm H}/r_{\rm d}=c/H(z)r_{\rm d}$, in terms of the sonic radius $r_{\rm d}$.

As we have highlighted throughtout this paper, the Alcock-Paczy\'nski diagnostic 
is unique (and the cleanest) because it is simply proportional to $(d_{\rm com}/r_{\rm d})
/(d_{\rm H}/r_{\rm d})$. Thus, $H_0$ and $r_{\rm d}$ completely cancel out, so one does 
not need to deal with these imprecisely known parameters. Other combinations of these 
distances depend to some degree on the ratio $Q\equiv c/H_0 r_{\rm d}$, which is
unknown. Nevertheless, by defining the quantity $x(z)$ in Equation~(\ref{eq:xz}),
one may still avoid using the model-dependent radius $r_{\rm d}$. Unfortunately, this
reduces the number of available data points, but the joint analysis using both
the AP and $D_v$ effects has produced some very useful results. 

In light of this work, we do not find direct evidence from the BAO peak 
measurements that the Universe is currently accelerating. The model favored by the AP 
test on its own is $R_{\rm h}=ct$, which has an expansion factor $a(t)\propto t$, 
suggesting that the cosmic expansion is proceeding at a constant rate. In fact, this 
model fits the BAO data better than {\it Planck}-$\Lambda$CDM and $\Lambda$CDM with 
an optimizable $\Omega_{\rm m}$, and it does this {\it without} the benefit of any 
free parameters. In addition, the value of $H_0$ is not required for any of this 
analysis, so its uncertain nature cannot be used as a possible reason for this disparity.

Interestingly, however, the model selection is reversed when the volume-averaged distance
measure is used on its own. Not surprisingly, one therefore finds that a joint analysis
using both the AP and $D_v$ diagnostics yields inconclusive evidence, favoring neither
model over the other.

Nevertheless, it is useful to point out from Figure~\ref{fig1} that the datum most important 
to this outcome is that corresponding to the Ly-$\alpha$ measurement at $\langle z\rangle =2.33$, 
which is arguably the most precise of the set we have used. The fact that this measurement is 
in tension with the standard model has been noted before\cite{Delubac:2015,Bautista:2017}, 
though the final SDSS-IV eBOSS release has cemented this disparity more emphatically. The 
prediction of $y(2.33)=1.96$ by {\it Planck}-$\Lambda$CDM differs from the measured value 
by over $2.2\sigma$. This inconsistency has been reconsidered following each successive 
enhancement of the SDSS quasar catalog, and is now established rather solidly. It provides 
an important affirmation of the turnover in $y(z)$ expected towards higher redshifts in 
$R_{\rm h}=ct$, which is completely lacking in $\Lambda$CDM (see fig.~\ref{fig1}).

Our conclusion that the BAO peak measurements in the final SDSS-IV release, 
together with the Alcock-Paczy\'nski and $D_v$ tests, do not favor either the $R_{\rm h}=ct$ 
universe or $\Lambda$CDM-$w$CDM, suggests that the BAO observations, on their own, cannot
be used to argue for an accelerating Universe. This conclusion somewhat confirms the outcome 
of many other comparative tests based on a broad range of observations. A recent compilation
of these results may be found in Table~2 of ref.~\cite{Melia:2018}. The impact of the work 
reported in this paper is being explored elsewhere, particulary with regard to the growing
likelihood that inflation may not have worked as expected, and may in fact have never
happened\cite{Melia:2013,Liu:2020}.

\section*{Acknowledgments}
We are very grateful to the anonymous referee for their thoughtful and helpful
review, which has led to several notable improvements to the contents of this manuscript.
F.M. is also grateful to Amherst College for its support through a John Woodruff Simpson Lectureship.
MLC acknowledges support from the Spanish Ministry of Economy and Competitiveness (MINECO)
under the grant PGC-2018-102249-B-100.

\section*{Appendix}
We may gauge the validity of the approximate form of $\chi^2$ in Equation~(13) by comparing
its outcome to that of a more detailed approach taking the covariance matrix, $\mathbb{C}$,
into account. As noted earlier, the covariance matrix for $d_{\rm com}$ and $d_{\rm H}$ in 
the binnings at $z=0.38$ and $0.51$ is available 
online.\footnote{https://svn.sdss.org/public/data/eboss/DR16cosmo/tags/v1-0-1/likelihoods/BAO-only/}
In this case, we would write
\begin{equation}
\chi^2=\mathbb{A}^T\mathbb{C}^{-1}\mathbb{A}\;,\label{eq:chiACA}
\end{equation}
where
\begin{equation}
\mathbb{A}\equiv
\left( \begin{array}{c}
              |y_{\rm th}(z_1)-y(z_1)| \\
              ...  \\
              |y_{\rm th}(z_{N_b})-y(z_{N_b})|
       \end{array} \right)\;,\label{eq:A}
\end{equation}
with $N_b$ the number of bins (with some correlation between adjacent pairs only), and
\begin{eqnarray}
\mathbb{C}_{ij}&\equiv& \langle[y_{\rm th}(z_i)-y(z_i)]\,[y_{\rm th}(z_j)-y(z_j)]\rangle \\
&=&{1\over N_{ij}-1}\sum_{k=1}^{N_{c,\,ij}}\left[y_{\rm th}(z_k)-y(z_k)\right]^2\;,
\end{eqnarray}
where $N_{ij}$ is the total number of galaxies in bins $i$ and $j$, and $N_{c,\,ij}$
is the corresponding number of galaxies common to these two bins. Thus,
\begin{equation}
\mathbb{C}_{ij}=\sigma_i\,\sigma_j{N_{c,\,ij}\over N_{ij}-1}\;.\label{eq:Cij}
\end{equation}

\vskip 0.2in
\noindent {\footnotesize{\bf Table 5.} Model comparison based solely on
the AP effect (i) using Equation~(\ref{eq:chiACA}), including the covariance matrix; (ii) 
optimized parameters (if any); (iii) AIC; and (iv) KIC.} 
\begin{table*}
\begin{center}
\begin{tabular}{lccrr}
Model & $\chi^2_{\rm dof}$ & Parameter & AIC$\;$ & KIC$\;$ \\ \hline\hline
\\
$\Lambda$CDM & 2.18 & $\Omega_{\rm m}=0.276\pm0.041$ & 
8.53 & 9.53 \\
{\it Planck}-$\Lambda$CDM\qquad\qquad & 1.98 & -- & 7.92 & 
7.92 \\
$R_{\rm h}=ct$ & 1.43 & -- & 5.70 & 5.70 \\ \\
\hline\hline
\end{tabular}
\end{center}
\end{table*}
\vfill\newpage
\vskip 0.2in
\noindent {\footnotesize{\bf Table 6.} Head-to-head model comparisons based solely
on the AP effect, using the AIC and KIC values in Table~5.}
\vskip -0.1in
\begin{table*}
\begin{center}
\begin{tabular}{lcccc}
Model Comparison & $\Delta {\rm AIC}$ & AIC & $\Delta {\rm KIC}$ & KIC \\
\hline\hline
\\
{\it Planck}-$\Lambda$CDM vs. $R_{\rm h}=ct$ & 2.22 & 
$25\%$ vs. $75\%$ & 2.22 & $25\%$ vs. $75\%$ \\
$\Lambda$CDM vs. $R_{\rm h}=ct$ & 2.83 & $20\%$ vs. $80\%$ & 
3.83 & $13\%$ vs. $87\%$ \\ \\
\hline\hline
\end{tabular}
\end{center}
\end{table*}

We define the relative number of common galaxies within the overlapping
bins to be proportional to $g_{ij}$. Therefore, 
\begin{equation}
\mathbb{C}_{ij}=\sigma_i\,\sigma_j\,g_{ij}\;.\label{eq:Cov}
\end{equation}
From Equations~(\ref{eq:chiACA}), (\ref{eq:A}) and (\ref{eq:Cov}), one may therefore write
\begin{eqnarray}
\chi^2&=&\sum_{i=1}^{N_b-1} k_i (1-g_{i,i+1}^2)\left({[y_{\rm th}(z_i)-y(z_i)]^2\over\sigma_i^2}
+\right.\nonumber \\
&\null&\qquad \left.{[y_{\rm th}(z_{i+1})-y(z_{i+1})]^2\over\sigma_{i+1}^2}-\right.\nonumber \\
&\null&\left. 2g_{i,i+1}
{|y_{\rm th}(z_i)-y(z_i)|\over \sigma_i}{|y_{\rm th}(z_{i+1})-y(z_{i+1})|\over 
\sigma_{i+1}}\right).\qquad
\end{eqnarray}
In this expression, $k_i=1$ for $i=1$ and $N_b-1$, while $k_i=1/2$ for all the others. For the
analysis in this paper (see Table~1), we also have $N_b=5$ and, 
taking $g_{12}=C_{12}/\sqrt{C_{11}\times C_{22}}$ from the published 
covariance tables for the DR12 BAO-only,\footnote{https://svn.sdss.org/public/data/eboss/DR16cosmo/tags/v1-0-0/likelihoods/BAO-only/ } for either $d_{\rm com}$ or $d_{\rm H}$ (which will also be the same 
for $y(z)$ because, as shown by Eq.~\ref{eq:Cij}, $g_{12}$ for any variable need only include
the number of galaxies that are common in the two bins), we also have
\begin{eqnarray}
g_{12}&=&0.42\nonumber \\
g_{23}&=&0\nonumber \\
g_{34}&=&0\nonumber \\
g_{45}&=&0\;.
\end{eqnarray}
The first three values reflect the overlap in redshift of the BOSS
and e-BOSS bins of galaxies, while the last two values are zero because
we neglect the correlations of galaxies with QSOs and of QSOs with the
Ly-$\alpha$ forest.

If we repeat the calculations in \S~2.1, though now using Equation~(\ref{eq:chiACA})
instead of Equation~(\ref{eq:chi2}), we find the outcomes and likelihoods summarized
in Tables~5 and 6. The numbers change slightly, but the relative
likelihoods essentially remain intact. Most importanty, the prioritization of models
based on the Alcock-Paczy\'nski effect in the BAO measurements shows that $R_{\rm h}=ct$
is favored over both {\it Planck} $\Lambda$CDM and a more generic version of the standard
model with $\Omega_{\rm m}$ as a free parameter.

\end{document}